\documentclass[runningheads]{svmult}

\usepackage{makeidx}   % allows index generation
\usepackage{graphicx}  % standard LaTeX graphics tool
\usepackage{subeqnar}  % subnumbers individual equations
\usepackage{multicol}  % used for the two-column index
\usepackage{physprbb}  % modified textarea for proceedings,
\makeindex             % used for the subject index
\begin{document}
\title*{HST Imaging of a $z=1.55$ Old Galaxy Group}
\toctitle{HST Imaging of a $z=1.55$ Old Galaxy Group}
\titlerunning{HST Imaging of a $z=1.55$ Old Galaxy Group}
\date{October 25, 2001}
\author{Andrew Bunker\inst{1,2} \and Hyron Spinrad\inst{2} \and Ross
McLure\inst{3,4} \and
Arjun Dey\inst{5} \and James Dunlop\inst{4} \and John Peacock\inst{4} \and Daniel
Stern\inst{2,6} \and
Rodger Thompson\inst{7} \and Ian Waddington\inst{8,9} \and Rogier Windhorst\inst{8}}
\authorrunning{Andrew Bunker et al.}

\institute{Institute of Astronomy, Madingley Road, Cambridge CB3~0HA, UK
\newline  {\tt email: bunker@ast.cam.ac.uk}
\and
Department of Astronomy, 601
Campbell Hall, Berkeley CA~94720, USA\and
Department of Astrophysics, Keble Road, Oxford OX1~3RH, UK\and
Institute for Astronomy, Blackford Hill, Edinburgh
EH9~3HJ, UK \and
KPNO/NOAO, 950 N.~Cherry Avenue, Tucson,
AZ~85726, USA \and
Jet Propulsion Laboratory/Caltech MS~169-327, Pasadena, CA~91109, USA \and
Steward Obs., University of Arizona, N.~Cherry Avenue,
Tucson AZ~85721, USA   \and
Department of Physics, Arizona State University, Tempe, AZ~85287, USA \and 
Department of Physics, Bristol University, Tyndall Avenue, Bristol,
BS8~1TL, UK}

\maketitle

\begin{abstract}
We present high-resolution imaging in the rest-frame optical of the weak
radio source LBDS53W091. Previous optical spectroscopy has shown that
this object has an evolved stellar population of age $>3$\,Gyr at
$z=1.55$, determined from the amplitude of rest-frame UV spectral
breaks. We have obtained deep Hubble Space Telescope imaging over 10
orbits with NICMOS camera 2, using the F160W $H$-band filter (1.6\,$\mu$m)
which is a good approximation to the rest-frame $R$-band.  Our
observations reveal a radial light profile which is well fit by a de
Vaucouleurs' $r^{1/4}$ law, with a scale length of $r_{e}=0.3''$
($3\,h_{50}^{-1}$\,kpc for $\Omega_{M}=0.3$ and
$\Omega_{\Lambda}=0$). The elliptical morphology of the radio galaxy
indicates a dynamically-evolved old system, consistent with the
spectroscopic results. Some surrounding objects lie on the
color:magnitude relation for a cluster at $z=1.55$, and are likely to be
associated.  The group of galaxies are somewhat more luminous than the
fundamental plane of ellipticals at $z=0$ subject to $(1+z)^{4}$ surface
brightness dimming, but are consistent with estimates of the fundamental
plane at high redshift subject to passive luminosity evolution from a
formation epoch of $z>3$.
\index{abstract}
\end{abstract}

%1) 
\section{INTRODUCTION}

There has been a revolution in recent years in finding and studying
star-forming galaxies at high redshift. An orthogonal approach to
studying galaxy assembly is to focus on old stellar populations which
formed at even earlier cosmic times. These old galaxies at intermediate
redshift ($z\sim 1-2$) comprise some of the ``Extremely Red Objects'', a
population with $(R-K)>6$\,mag.  Here we present a morphological analysis of
one of the oldest galaxies known at high redshift. The weak 1-mJy radio
galaxy LBDS53W091 (1) has a stellar age of $>3$\,Gyr at $z=1.55$, based the
amplitude of rest-frame UV spectral breaks seen in Keck/LRIS spectra
(2,3) with a main-sequence turn-off around an F2-F6 star.  If this
spectral age-dating is correct, then the galaxy should be a
dynamically-evolved elliptical. Ellipticals at the current epoch are
remarkably homogeneous, with tight scaling relations - the Fundamental
Plane (4,5). It is important to understand the true nature of this
apparently old galaxy in the early Universe, and so trace the evolution
of these scaling relations. Therefore, we have obtained deep,
high-resolution imaging with the Hubble Space Telescope (HST) in the
near-infrared and optical, allowing us to study its morphology and
determine the structural parameters.

\begin{figure}
\centering
\resizebox{0.9\textwidth}{!}{\includegraphics{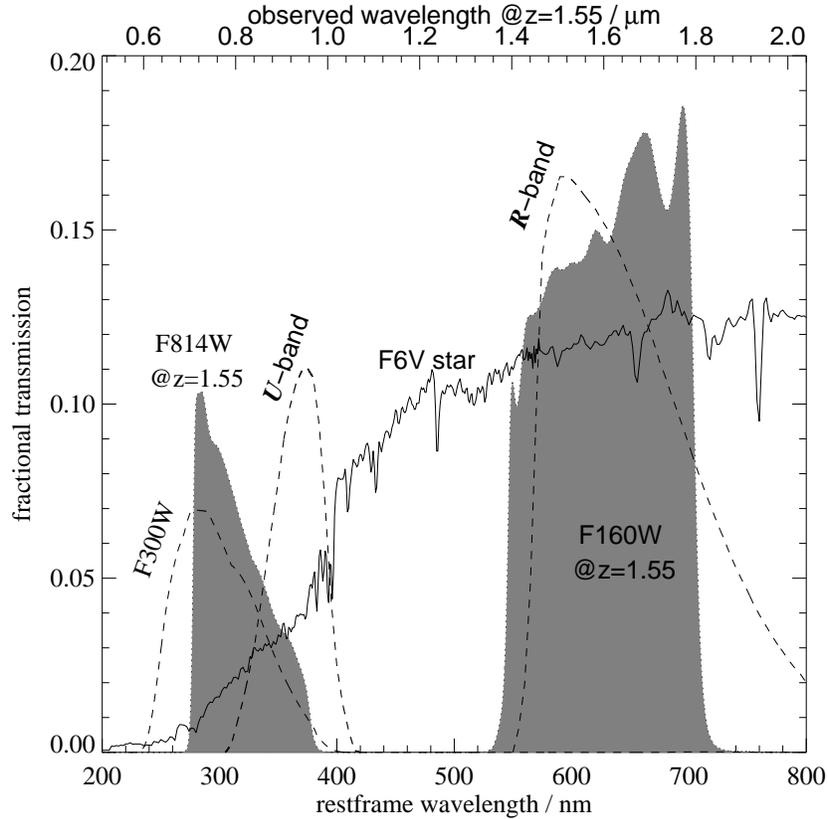}}
%Figure 1
\caption{HST $I$- and $H$-bands (shaded), showing location in the
rest-frame of a galaxy at $z=1.55$ with the spectral type of LBDS53W091
(solid line)}
\end{figure}

%2) 
\section{OBSERVATIONS WITH HST}

We observed LBDS53W091 with HST over 10 orbits (19\,ksec) using the Near
Infrared Camera/Multi-Object Spectrograph, NICMOS (6). We used the
diffraction-limited camera NIC2 (0.076\,arcsec/pixel and a 19\,arcsec field)
and the F160W $H$-band filter (1.6\,$\mu$m) which is a good approximation
to the rest-frame $R$-band (see figure~1). Our near-infrared observations
were complemented by a 2 orbit (5\,ksec) WFPC2 image using the F814W
``$I$-band'' filter (0.8\,$\mu$m), which at $z=1.55$ is a good match to the
rest-frame F300W filter. We used ``drizzle'' (7) to improve the resolution
through sub-pixel sampling. Figure~2 shows the registered $I$- and $H$-band
images. Prominent in the near-infrared is a group of extremely red
galaxies around the location of the radio source.

\begin{figure}
\resizebox{\textwidth}{!}{\includegraphics{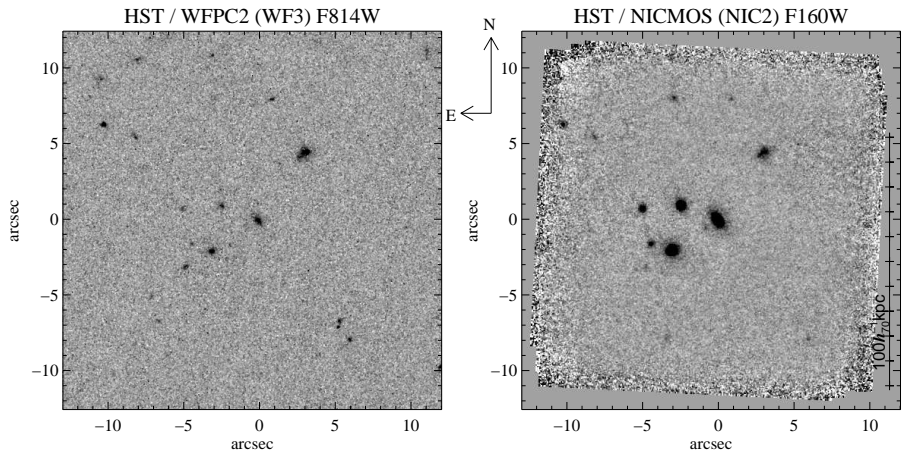}}
%Figure 2
\caption{The registered $I$- and $H$-band
images. Prominent in the near-infrared is a group of extremely red
galaxies around the location of the radio source (the coordinate
zero-point).}
\end{figure}

%3) 
\section{SURFACE BRIGHTNESS PROFILE}

A range of two-dimensional models were convolved with the point spread
function, then compared against the data to determine the best-fit
parameters (figure~3). We fit the radial dependence of the surface
brightness profile with various functional forms: an $r^{1/4}$ bulge
(8); an exponential disk; a generic S\'{e}rsic $r^{1/n}$ profile (9); and a
composition of disk and bulge (10).  Our observations reveal a radial
light profile which is well fit by a de Vaucouleurs' $r^{1/4}$ law, with
a scale length of $r_{e}=0.3$\,arcsec ($3\,h^{-1}_{50}$\,kpc for
$\Omega_{M}=0.3$ and $\Omega_{\Lambda}=0$).  The optical $I$-band
morphology (sampling the rest-UV and sensitive to younger, bluer stars)
does show some evidence for a disk component, but the near-infrared
$H$-band light is completely bulge-dominated.  The elliptical morphology
of the radio galaxy indicates a dynamically-evolved old system,
consistent with the spectroscopic results.  Down to our detection
limits, we find no evidence for an unresolved central point source (11)
due to AGN activity in this weak radio galaxy.

\begin{figure}
\resizebox{0.45\textwidth}{!}{\rotatebox{270}{\includegraphics{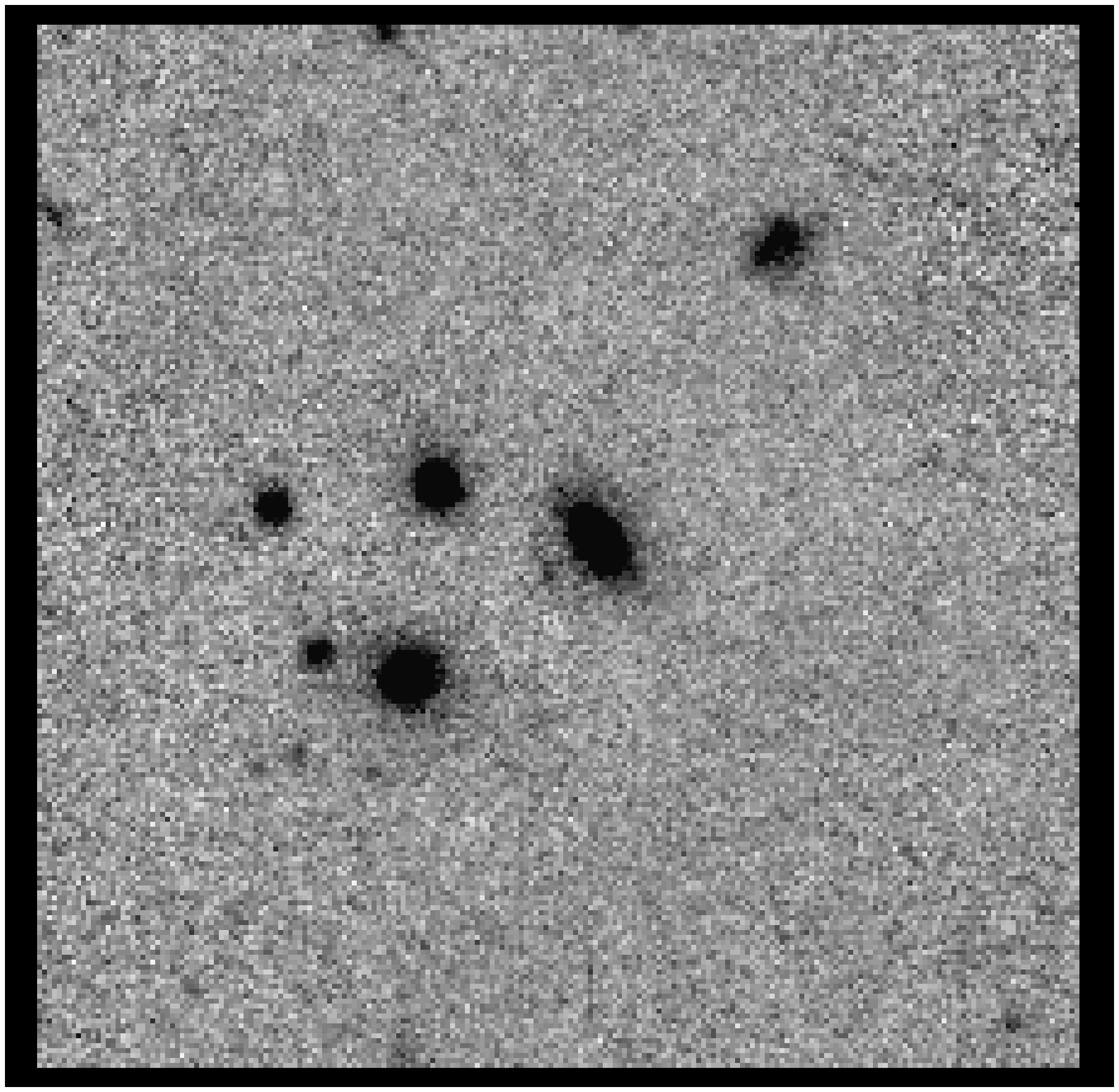}}}
\resizebox{0.45\textwidth}{!}{\rotatebox{270}{\includegraphics{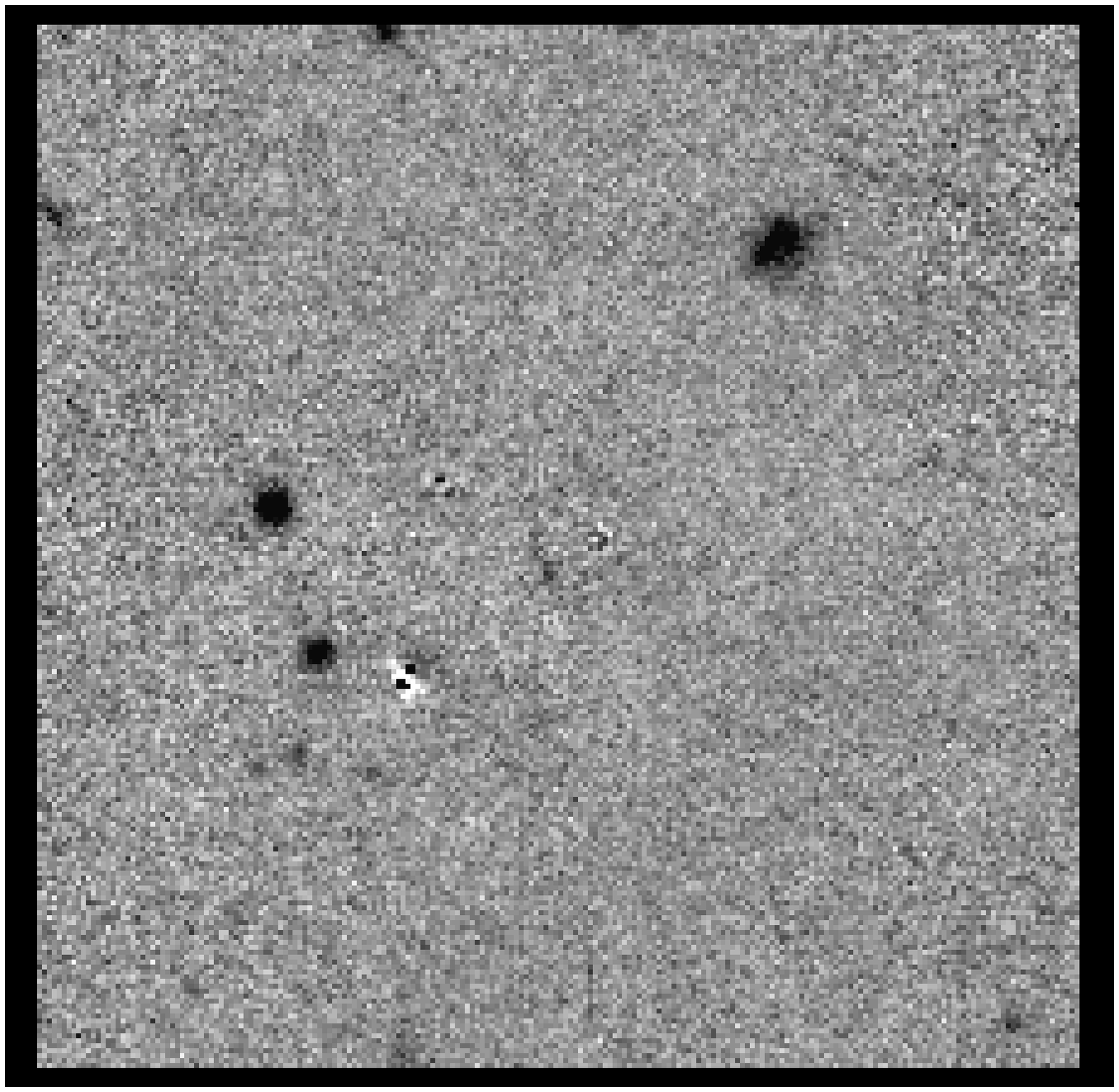}}}
%Figure 3
\caption{The NICMOS $H$-band image (left), and with the best-fit
two-dimensional de Vaucouleur $r^{1/4}$ models subtracted (right) for
the three brightest galaxies. The small residuals indicate that these
galaxies are bulge-dominated/elliptical in nature. The model fits have been 
produced using the GIM2D package of Simard et al.\ (1999).}
\end{figure}

%4) 
\section{THE FUNDAMENTAL PLANE AT $z=1.55$}

The ``Fundamental Plane'' (4,5) provides a tight correlation between the
scale length ($r_e$), surface brightness ($I_e$) and central velocity
dispersion ($\sigma_c$) of the form
$r_e=k\,\sigma_c^{1.2}\,I_e^{-0.83}$; the precise physical origin of the
fundamental plane is uncertain, but in essence it is a relation between
galaxy mass and mass-to-light ratio ($M/L\propto M^{1/4}$, see 12).
Recent work on clusters at moderate redshifts out to $z\approx 0.8$ has
suggested an evolution in the zero-point of the fundamental plane
(13,14), consistent with passive luminosity evolution of the elliptical
stellar population which had formed at much higher redshifts. It is
important to locate high redshift cluster ellipticals, such as those
associated with LBDS53W091, on the fundamental plane as the inferred
mass-to-light ratios may tell us how close these galaxies are to the
formation epoch of ellipticals.  In the case of LBDS53W091, no direct
measurement of the central velocity dispersions exists - although our
NICMOS imaging has yielded good determinations of the half-light radius
and mean surface brightness. Plotting the edge-on projection of the
fundamental plane for Coma (figure 4) shows that LBDS53W091 is more
luminous than the fundamental plane at $z\approx 0$, correcting only for
the $(1+z)^{4}$ surface brightness dimming. However, if we also correct
for passive luminosity evolution from a formation epoch of $z>3$ then
this $z=1.55$ galaxy would lie on the fundamental plane if
$\sigma\approx 300$\,km/s (comparable with indirect velocity dispersion
estimates -- see 15).

\begin{figure}
\centering
\resizebox{0.9\textwidth}{!}{\includegraphics{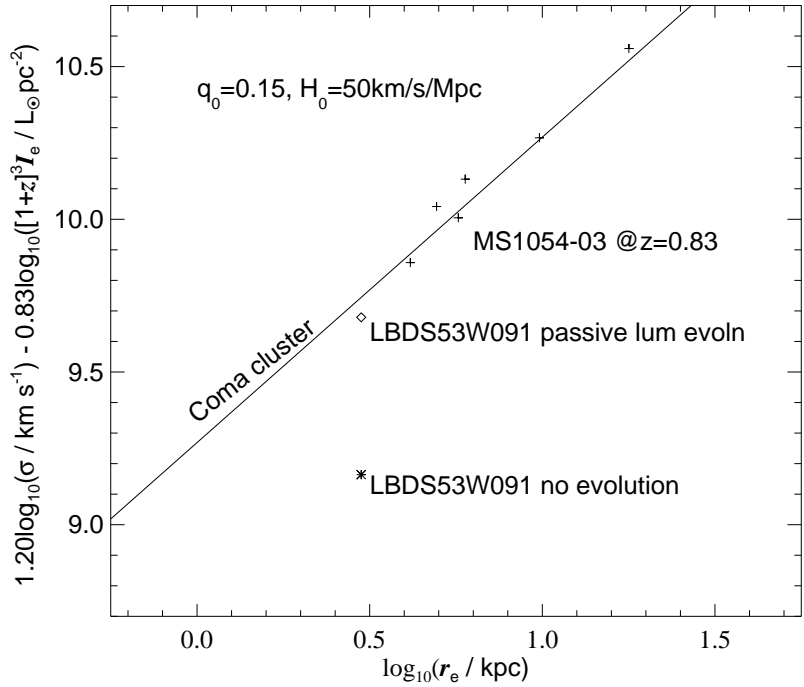}}
%Figure 4
\caption{
The Coma fundamental plane viewed edge on (solid line). The crosses are
ellipticals in the cluster MS1054-03 at $z=0.83$ (see 14), corrected for
passive luminosity evolution of $\Delta Mag\approx 0.4\times z$. A
cosmology of $\Omega_{M}=0.3$, $\Omega_{\Lambda}=0$,
$H_{0}=50$\,km/s/Mpc is used. The asterisk represents the galaxy
LBDS53W091, assuming a velocity dispersion of $\sigma\sim
300$\,km/s. This point lies well below the fundamental plane, implying
the galaxy is brighter than present-day ellipticals. However, once
correction for passive luminosity evolution since $z=1.55$ has been made
(diamond symbol), LBDS53W091 lies close to the fundamental plane for a
reasonable range of velocity dispersions.}
\end{figure}

%5)  
\section{AN OLD GROUP  AT $z=1.55$?}

Some surrounding objects lie on the color:magnitude relation for a
cluster at $z=1.55$, and are likely to be associated.  The tightness of
the color:magnitude diagram in clusters could be attributable either to
a high formation redshift, or an extremely synchronized, coeval
star-formation history of the member galaxies (16).  The color:magnitude
diagram determined at $z\approx 0$ from Coma is redder by
$\Delta(U-B)\sim 0.5$\,mag than
the observed envelope for the field of LBDS53W091 (figure 5). Once
again, this is entirely consistent with passive luminosity evolution and
a high formation redshift.

\begin{figure}
\resizebox{0.9\textwidth}{!}{\includegraphics{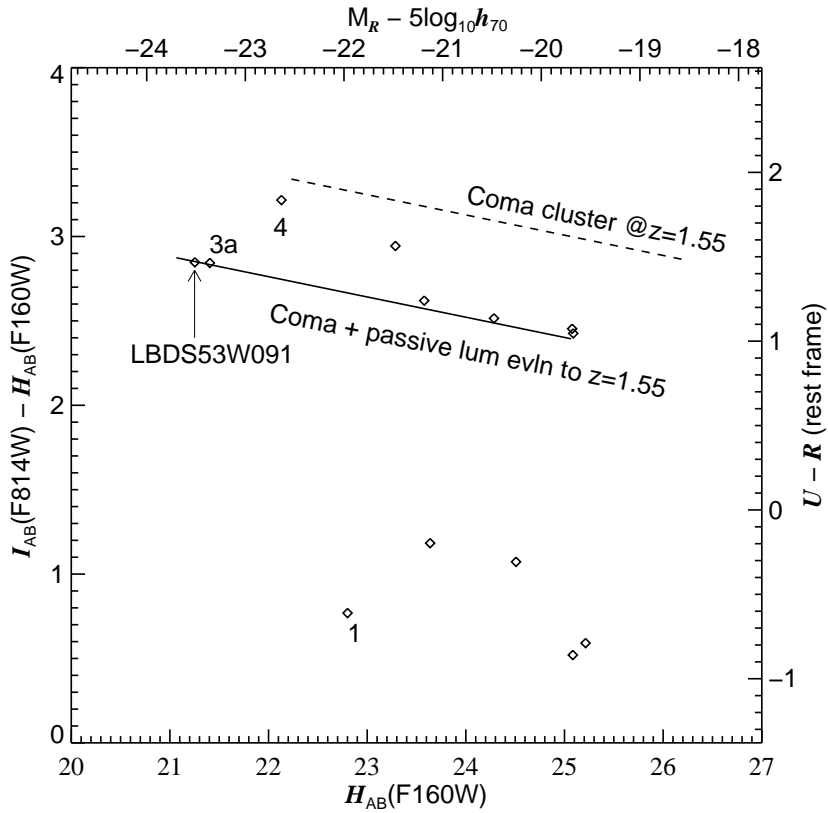}}
%Figure 5
\caption{The color:magnitude diagram for objects in the field of
LBDS53W091. Some 
are significantly more blue than the radio galaxy, and are presumably
foreground: ``galaxy 1'' has a spectroscopic
redshift of $z=1.01$ (see 3). However, there are half a dozen objects that lie close to
the color:magnitude relation from Coma, corrected for passive luminosity
evolution (solid line). This may be evidence for a cluster around
LBDS53W091 at $z=1.55$
-- indeed, ``galaxy 3a'' has the same spectroscopic redshift (see 3).}
\end{figure}

%INDEX%%%%%%%%%%%%%%%%%%%%%%%%%%%%%%%%%%%%%%%%%%%%%%%%%%%%%%%%%%%%%%%
\clearpage
\addcontentsline{toc}{section}{Index}
\flushbottom
\printindex
%%%%%%%%%%%%%%%%%%%%%%%%%%%%%%%%%%%%%%%%%%%%%%%%%%%%%%%%%%%%%%%%%%%%%

\end{document}